\documentclass[pmlr,twocolumn,10pt]{jmlr} 

\usepackage{enumitem}
\usepackage{xspace}
\usepackage{multirow}
\usepackage{adjustbox}
\usepackage{graphicx}
\usepackage{boldline}

\usepackage{booktabs}
\usepackage{mathtools}
\usepackage{amsfonts}
\usepackage{makecell}
\usepackage{amssymb}
\usepackage{nccmath}
\usepackage{bm,amsmath}
\usepackage{xspace}
\usepackage{color}
\usepackage[english]{babel}
\usepackage[utf8]{inputenc}
\usepackage{amsmath}
\usepackage{graphicx}

\usepackage[colorinlistoftodos]{todonotes}

\usepackage{url}            
\usepackage{booktabs}       
\usepackage{amsfonts}       
\usepackage{array,multirow}

\usepackage{subcaption}
\usepackage{wrapfig}

\usepackage{enumitem}

\newcommand{\eb}{\mathbf{e}}

\newcommand{\sbb}{\mathbf{s}}

\newcommand{\vb}{\mathbf{v}}
\newcommand{\wb}{\mathbf{w}}

\usepackage{siunitx}

\newcommand{\model}{ViewXGen\xspace}

\usepackage[switch]{lineno}

\theorembodyfont{\upshape}
\theoremheaderfont{\scshape}
\theorempostheader{:}
\theoremsep{\newline}

\jmlrvolume{LEAVE UNSET}
\jmlryear{2024}
\jmlrsubmitted{LEAVE UNSET}
\jmlrpublished{LEAVE UNSET}
\jmlrworkshop{Conference on Health, Inference, and Learning (CHIL) 2024} 

\title[Vision-Language Generative Model for View-Specific Chest X-ray Generation]{Vision-Language Generative Model for \titlebreak
View-Specific Chest X-ray Generation}

\author{%
\Name{Hyungyung Lee} \Email{ttumyche@kaist.ac.kr}\\
\addr KAIST, Republic of Korea
\AND
\Name{Da Young Lee\nametag{\thanks{Work done at KAIST}}}\Email{dyan.lee717@gmail.com}\\
\addr Deep-in-Sight Co., Republic of Korea
\AND
\Name{Wonjae Kim} \Email{wonjae.kim@navercorp.com}\\
\addr NAVER AI Lab, Republic of Korea
\AND
\Name{Jin-Hwa Kim} \Email{j1nhwa.kim@navercorp.com}\\
\addr NAVER AI Lab, Republic of Korea \\
\addr AI Institute of Seoul National University, Republic of Korea
\AND
\Name{Tackeun Kim} \Email{tackeun.kim@snu.ac.kr} \\
\Name{Jihang Kim} \Email{radio622@gmail.com} \\
\Name{Leonard Sunwoo} \Email{leonard.sunwoo@gmail.com} \\
\addr Seoul National University Bundang Hospital, Republic of Korea
\AND
\Name{Edward Choi} \Email{edwardchoi@kaist.ac.kr}\\
\addr KAIST, Republic of Korea
}

\begin{document}

\maketitle

\begin{abstract}
Synthetic medical data generation has opened up new possibilities in the healthcare domain, offering a powerful tool for simulating clinical scenarios, enhancing diagnostic and treatment quality, gaining granular medical knowledge, and accelerating the development of unbiased algorithms.
In this context, we present a novel approach called \model, designed to overcome the limitations of existing methods that rely on general domain pipelines using only radiology reports to generate frontal-view chest X-rays.
Our approach takes into consideration the diverse view positions found in the dataset, enabling the generation of chest X-rays with specific views, which marks a significant advancement in the field.
To achieve this, we introduce a set of specially designed tokens for each view position, tailoring the generation process to the user's preferences.
Furthermore, we leverage multi-view chest X-rays as input, incorporating valuable information from different views within the same study. This integration rectifies potential errors and contributes to faithfully capturing abnormal findings in chest X-ray generation.
To validate the effectiveness of our approach, we conducted statistical analyses, evaluating its performance in a clinical efficacy metric on the MIMIC-CXR dataset.
Also, human evaluation demonstrates the remarkable capabilities of \model, particularly in producing realistic view-specific X-rays that closely resemble the original images.
\end{abstract}

\paragraph*{Data and Code Availability}
We use the MIMIC-CXR dataset, which is available on the PhysioNet repository \citep{johnson2019mimic}.
Our implementation code is available at this repository\footnote{\url{https://github.com/ttumyche/UniXGen}}.

\paragraph*{Institutional Review Board (IRB)}
This research does not require IRB approval.

\section{Introduction}
\label{sec:intro}
\begin{figure}[h]
    \centering
    \includegraphics[width=0.9\columnwidth]{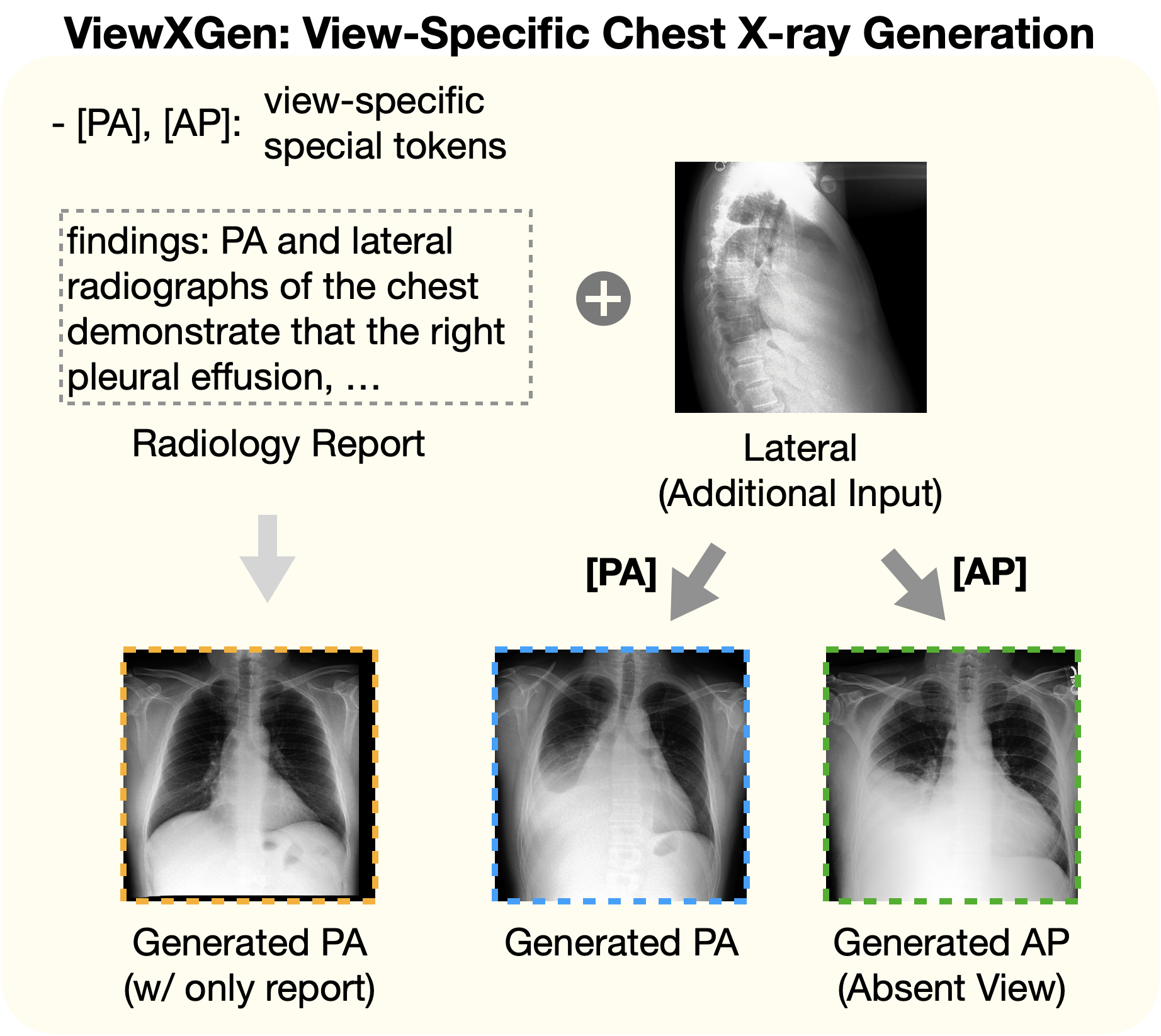}
    \caption{
    We introduce a view-specific chest X-ray generation model. \model leverages view-specific special tokens to empower its ability to capture unique features from different views. Additionally, the integration of multi-view chest X-rays as input enhances the overall generation quality.
    }
\label{fig:overview}
\end{figure}

Chest X-ray generation has become increasingly significant in the medical field, yet prior studies \citep{packhauser2022generation, chambon2022adapting, chambon2022roentgen} have notably missed two crucial aspects: First, there's a heavy reliance on radiology reports for generating chest X-rays, which disregards the rich information available in other X-ray views within the same study. Second, the importance of controlling view positions has been neglected, despite the fact that various views reveal diverse characteristics due to the angle of the X-ray beam \citep{puddy2007interpretation}. 

To address these, we introduce \model, a versatile generative model tailored for generating synthetic chest X-rays that are specific to view, symptom, and patient.
Figs.~\ref{fig:gen_cxr}, and~\ref{fig:retrieval} showcase detailed examples of the generated chest X-rays by our model, demonstrating its capabilities.
This approach sets our work apart from earlier studies, showcasing a wide range of clinical applications:
1) Filling in Missing Data: Our model can address gaps by generating specific views that may have been mentioned in a report but are currently missing.
Moreover, it enriches the generated images with patient information observed in other views, including gender, age, and obesity level.
Upon investigating the presence of missing data in MIMIC-CXR \citep{johnson2019mimic}, it was discovered that among 27,859 studies where specific views were explicitly mentioned in the reports, 1,565 of these studies (5.62\%) did not contain the mentioned views.
2) Reducing the Need for Additional Imaging: Our model provides a solution for scenarios where obtaining certain views is impractical due to patient conditions or limitations in medical equipment. By generating the necessary views, it conserves both time and resources, offering a way to acquire patient-specific images without additional imaging.
3) Enhancing Education and Training: The ability to create and analyze customized views and patient cases empowers medical students and professionals. This feature aids in deepening the understanding of how various conditions manifest across different X-ray views, thereby improving diagnostic capabilities and expanding anatomical knowledge.
4) Augmenting Data for Rare Conditions: Our model excels in generating images for a wide range of scenarios, including plausible yet rare conditions, enriching datasets with unique views that spotlight uncommon pathologies and aiding in the research and diagnosis of rare conditions.

To achieve these, we introduce a set of special tokens tailored to each view position, including posterior-anterior (PA), anterior-posterior (AP), and lateral views, and employ a simplified architectural design by combining VQ-GAN \citep{esser2021taming} and Performer \citep{choromanski2020rethinking}, which is an efficient Transformer-based framework. Specifically, we utilize VQ-GAN as an image tokenizer, enabling the conversion of chest X-ray images into sequences of discrete tokens.
The adoption of Performer enhances computational and memory efficiency, crucial for processing long paragraph reports and high-resolution multi-view chest X-rays that result in long-range sequences.
By leveraging this approach, our model demonstrates the capability to handle diverse input formats, ranging from single to multi-view images. 

We evaluate our model on MIMIC-CXR \citep{johnson2019mimic}.
The experimental results show that \model achieves better performance on both standard metrics such as FID \citep{huang2017densely} and clinical efficacy metrics such as 14-diagnosis classification over several baselines.
Furthermore, human evaluation shows that \model can generate realistic chest X-rays comparable to the original image, and the view-specific special tokens capture the refined features of each view, encouraging the model to generate appropriate view-specific X-rays.

Our contributions can be summarized as follows.
\begin{enumerate}[label=\arabic*), noitemsep, nolistsep]
    \item \textbf{Pioneering Approach}: Our work marks the first attempt to generate view-specific chest X-ray images with multimodal input in the medical domain. Additionally, we introduce special tokens that are simple yet effective for generating specific view positions. These tokens provide precise control over the view generation process, enabling our model to produce X-rays from various view positions.
    \item \textbf{Novel Task}: We propose a novel task of generating chest X-rays with specific views, such as PA, AP, and Lateral views. This task addresses the limitations of previous approaches that primarily focused on generating frontal views and disregarded the multi-view nature of the dataset.
    \item \textbf{Multi-View Integration}: By leveraging multi-view chest X-rays, our model demonstrates the potential to generate more accurate chest X-rays that capture abnormal findings and patient characteristics present in additional X-rays. This integration of multi-view information improves the fidelity and diagnostic quality of the generated chest X-rays.
\end{enumerate}

\section{Related Works}
\subsection{Chest X-ray Generation}
With the growing demand to access high quality medical data and the success of generative models such as GANs \citep{goodfellow2020generative}, and diffusion models \citep{ho2020denoising}, chest X-ray generation has gained a lot of attention.
Chambon et al. \citep{chambon2022adapting} and Packhauser et al. \citep{packhauser2022generation} adopt a latent diffusion model \citep{rombach2022high} for class-conditional generation.
However, these works only focus on specific diseases and do not utilize radiology reports that contain rich medical domain knowledge.
Recently, Chambon et al. \citep{chambon2022roentgen} have taken advantage of radiology reports for conditional generation, but they only use the impression section of the reports.
Furthermore, they cannot generate view-specific chest X-rays or accept multiple views as input.

\subsection{Image Tokenization}
Many efforts have been made to convert images into discrete tokens like natural language, as this provides a compact and efficient representation compared to using raw pixels.
Based on the success of VQ-VAE \citep{van2017neural},
Esser et al. \citep{esser2021taming} introduced VQ-GAN with a discriminator and a perceptual loss for high-resolution images.
Recently, diffusion models have achieved promising performance in generating high-quality samples in continuous domains (\textit{e.g.}, image \citep{ramesh2022hierarchical} and audio \citep{saharia2022photorealistic}).
However, the models are not flexible to take arbitrary input from single to multiple images.

\subsection{Efficient Transformer}
Transformer \citep{vaswani2017attention} has proven to be highly adaptable to both vision and language tasks with its task-agnostic design and generalization capabilities.
However, the self-attention mechanism increases the computational and memory cost quadratically by the input sequence length.
As we utilize long paragraph reports and high resolution multi-view chest X-rays, we adopt Performer \citep{choromanski2020rethinking}, an efficient Transformer-based model to reduce the quadratic complexity to linear.
They approximate the standard Transformer attention using positive orthogonal random features to kernelize the softmax operation.

\section{Method}
\begin{figure*}[h]
    \vspace{-10mm}
    \centering
    \includegraphics[width=\textwidth]{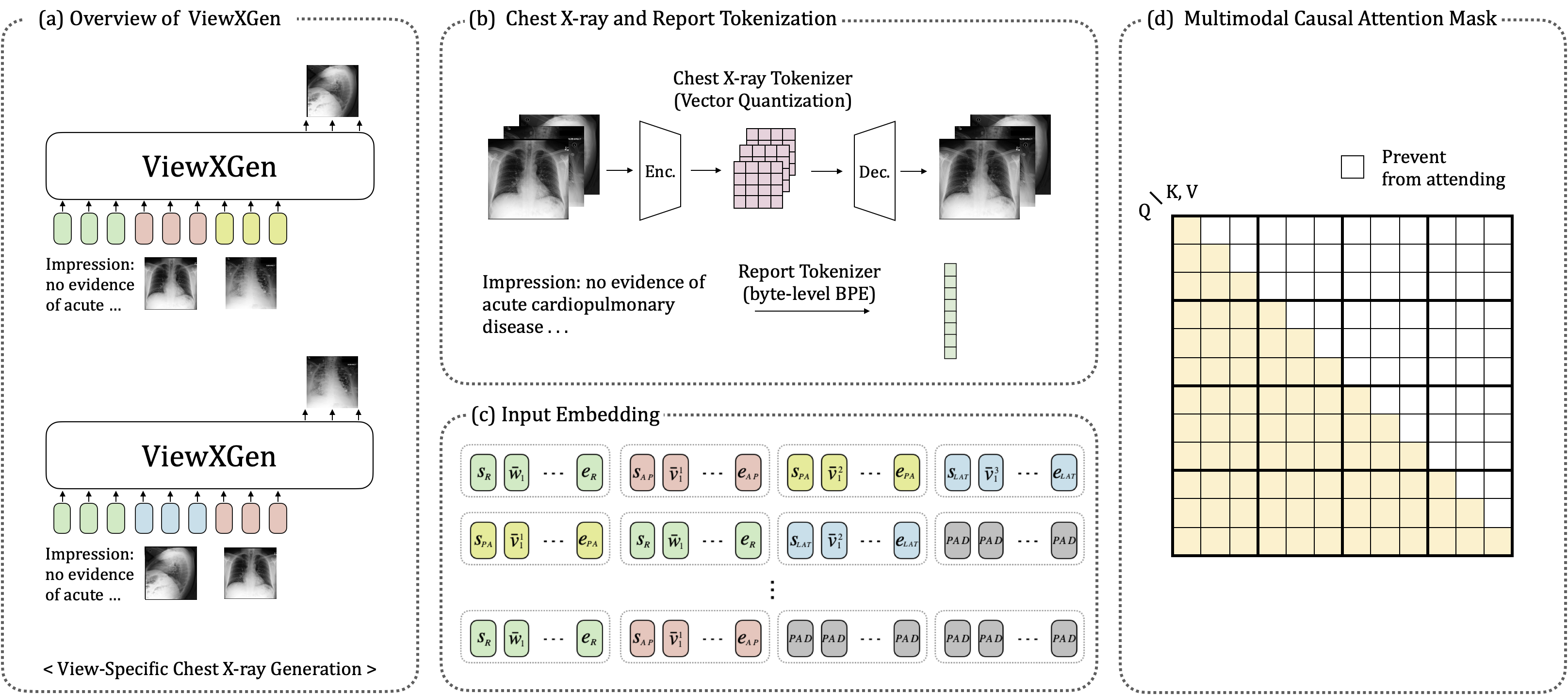}
    \caption{
    Overview of \model architecture.
    (a) \model is designed to generate chest X-rays with specific views, such as AP, PA, and Lateral views.
    (b) Images are tokenized via VQ-GAN, and reports are tokenized via a byte-level BPE tokenizer.
    (c) A minibatch consists of input sequences consisting of AP/PA/Lateral X-rays and a report in random order.
    (d) We use a causal attention mask to simultaneously handle multi-view X-rays and a report.
    }
\label{fig:architecture}
\end{figure*}

Fig.~\ref{fig:architecture} shows the overall depiction of \model.
Notably,
1) \model leverages a series of chest X-rays and a corresponding report from the same study as input, enhancing the quality of the generated chest X-rays.
2) To enable precise control over the generation of chest X-rays with specific views, we integrate special tokens tailored to each view type.

\subsection{Input Embedding}
\label{sec:method_input_embedding}
\subsubsection{Image Tokenization}
We first train VQ-GAN \citep{esser2021taming} to encode chest X-rays into a discrete latent space, enabling us to represent each image as a sequence of discrete tokens.
This model consists of an encoder $E$, a decoder $G$, and a fixed-size learnable codebook $C = \{e_m\}^M_{m=1}$ of size $M$, where $e_m \in \mathbb{R}^n$.
Given an image $\mathbf{x} \in \mathbb{R}^{H \times W \times 3}$, the encoder encodes the input image into a continuous feature map $\mathbf{z} = E(\mathbf{x}) \in \mathbb{R}^{h \times w \times n}$.
Then, we obtain a quantized feature map $\hat{\mathbf{z}} \in \mathbb{R}^{h \times w \times n}$ and its sequence of visual tokens $\{v_1, \ldots, v_{h \times w}\}$, \textit{a.k.a.}, discrete codes as follows:
\begin{equation*}
     \hat{\mathbf{z}}_{ij} = Q (\mathbf{z}_{ij}) = e_m
     , \quad m = \arg \min_{k} \|\mathbf{z}_{ij} - e_k\| = v_{ij}
\end{equation*}
where $Q(\cdot)$ denotes an element-wise quantization operation that performs the nearest neighbor search, $\mathbf{z}_{ij} \in \mathbb{R}^n$ is a feature vector at $(i, j)$, and $v_{ij}$ is its code.
The decoder then maps the quantized feature map back to the original input $\hat{\mathbf{x}} = G(\hat{\mathbf{z}}) \in \mathbb{R}^{H \times W \times 3}$.

The encoder-decoder model and codebook are optimized using the following objectives:
\begin{equation*}
        L_{VQ}(E, G, C) = \|\mathbf{x} - \hat{\mathbf{x}}\|^2_2 +
        \|sg[\mathbf{z}] - \hat{\mathbf{z}}\|^2_2
        + \beta \|sg[\hat{\mathbf{z}}] - \mathbf{z}\|^2_2
\end{equation*}
where the first term is a reconstruction loss, the second term optimizes the codebook embedding, the last term refers to a commitment loss with weighting factor $\beta$, and $sg$ refers to a stop-gradient.
To further enhance the reconstruction quality, VQ-GAN incorporates a discriminator $D$ and perceptual loss as follows:
\begin{equation*}
        L_{GAN}(\{E, G, C\}, D) = [\log D(\mathbf{x}) + \log (1 - D(\hat{\mathbf{x}})]
\end{equation*}
Finally, the model is optimized as follows:
\begin{equation*}
    L_{VQGAN} = L_{VQ}(E, G, C) + \lambda L_{GAN}(\{E,G,C\}, D)
\end{equation*}
where $\lambda$ is an adaptive weight.
This method allows the model to learn a compact and discrete representation of the images.

\subsubsection{Chest X-ray Embedding}
Using the image tokenizer described above, chest X-rays of multiple views from the same study are individually tokenized into a sequence of discrete visual tokens, surrounded by special tokens to differentiate between different views, \textit{e.g.} $\{[SOS_{PA}], v_1, \ldots, v_{h \times w}, [EOS_{PA}]\}$ for a PA-view X-ray.
Additionally, if the study has fewer images than $k$\footnote{In our work, we use $k=3$ to include PA, AP, and Lateral view.}, we add padding tokens to ensure that all input sequences have the same length.
For example, the final embeddings of a PA-view X-ray is $\vb_{PA} = \{ \sbb_{PA}, \bar{\vb}_1, \ldots, \bar{\vb}_{h \times w}, \eb_{PA} \}$, where $\sbb_{PA}, \eb_{PA} \in \mathbb{R}^{d}$ respectively denote the embeddings of the special tokens, $\bar{\vb}_{i} \in \mathbb{R}^{d}$ 
is acquired by summing the visual embedding and axial positional embedding \citep{ho2019axial, kitaev2020reformer}:
\begin{equation*}
    \bar{\vb}_i =  f_{VE}(v_i) + f_{VP}(i)
\end{equation*}
where $f_{VE}(\cdot)$ and $f_{VP}(\cdot)$ are the visual embedding and axial positional embedding functions, respectively.

\subsubsection{Radiology Report Embedding}
We first split a report into word tokens with a byte-level BPE tokenizer \citep{wang2020neural} and surround them with special tokens, \textit{e.g.} $\{ [SOS], w_1, \ldots, w_T, [EOS] \}$.
The final embeddings for the report is $\wb=\{\sbb_{R}, \bar{\wb}_1, ..., \bar{\wb}_T, \eb_{R}\}$, where $\sbb_{R}, \eb_{R} \in \mathbb{R}^{d}$ respectively denote the embeddings of the special tokens, $\bar{\wb}_{i} \in \mathbb{R}^{d}$ is obtained by summing up
the word embedding and sinusoid positional embedding:
\begin{equation*}
    \bar{\wb}_i =  f_{WE}(w_i) + f_{WP}(i)
\end{equation*}
where $f_{WE}(\cdot)$ and $f_{WP}(\cdot)$ are the word embedding and sinusoidal positional embedding functions, respectively.

\subsection{Multi-view Chest X-ray Generative Model}
We design a model for multi-view chest X-ray generation by treating the task as a sequence generation task.
Incorporating the Transformer architecture \citep{vaswani2017attention}, our model is trained with a multimodal causal attention mask, which is designed to handle multimodal input while still maintaining the causal constraints of the standard causal mask as shown in Fig.~\ref{fig:architecture} (d).
The attention mask $M \in R^{S\times S}$ can be represented as follows:
\begin{equation*}
    M_{ij} \,\text{=}\, 
        \begin{cases}
            0, \quad \enskip \text{if i $\leq$ j} \\
            \mathbin{-}\infty, \enskip \text{otherwise} \\
        \end{cases}
     \quad i, j \,\text{=}\, 1, ..., S.
\end{equation*}
where a value of 0 indicates allow to attend, while $\mathbin{-}\infty$ prevents from attending, and $S = k \times (h\times w + 2) + T + 2$.
This attention mechanism differs from the sequence-to-sequence attention mask \citep{dong2019unified} as it treats all modalities as targets for generation, allowing the model to simultaneously learn each modality conditioned on the preceding modalities along with the first modality which performs unconditional generation in each iteration.

The conventional self-attention mechanism is widely recognized for its expressive capabilities:
\begin{equation*}
\begin{aligned}
Attention(Q, K, V) & = \operatorname{softmax}\left(\frac{Q K^{T}}{\sqrt{d_{k}}}+M\right) V \\
                   & = AV
\end{aligned}
\end{equation*}
where $Q$, $K$, $V$, and $d_{k}$ indicate queries, keys, values, and dimensions of queries and keys, respectively. However, when aiming for scalability and addressing long-range sequences, its computational demands can become a bottleneck.

To handle this with limited resources, we adopt the Performer \citep{choromanski2020rethinking} as an alternative that enhances computational efficiency.
Following Performer, we utilize the FAVOR+ algorithm which uses positive orthogonal random features to approximate the softmax function with linear space and time complexity, allowing the model to compute the attention score more efficiently and reduce memory consumption.
For causal attention, we also adopt a prefix-sum mechanism to avoid storing an explicit lower-triangular regular attention matrix.
The mechanism of the FAVOR+ algorithm for unidirectional attention are delineated below:
\begin{itemize}[leftmargin=*]
    \item Outer Product Computation: For each key $k_i$ and value $v_i$, compute the outer product using random features designated for keys:
\begin{equation*}
\Phi_k(k_i) v_i^T
\end{equation*}
where $\Phi_k$ stands for the random features corresponding to key.

    \item Prefix-Sum Matrix Update: Iteratively accumulate the outer products to update the prefix-sum matrix:
\begin{equation*}
P_i = P_{i-1} + \Phi_k(k_i) v_i^T
\end{equation*}
Notably, $P_0$ starts initialized to zero.

    \item Attention Matrix Row Generation: For every iteration, the most recent prefix-sum is multiplied with the random feature vector pertaining to a query. This yields a new row for the $AV$ matrix:
\begin{equation*}
AV_i = \Phi_q(q_i) P_i
\end{equation*}
\end{itemize}

where $\Phi_q$ stands for the random features corresponding to query.

To encapsulate the operation in matrix terminology:
\begin{equation*}
AV_i = \Phi_q(q_i) \sum_{j=1}^{i} \Phi_k(k_j) v_j^T
\end{equation*}
with $AV$ signifying the matrix generated by the attention mechanism.

During training, we concatenate a series of chest X-rays and report embeddings from the same study in random order to form a single input sequence as shown in Figure~\ref{fig:architecture} (c), which is then fed into the model.
\model is trained to minimize the negative log-likelihood of the next token given the previous tokens.
Given $[\wb;\vb^1;...;\vb^k]$ as the input sequence, for example, the loss function is formulated as follows:
\small
\begin{equation*}
    \begin{split}
    L = & \sum_{i=1}^{n} - logP(w_{i}|w_{0:i-1}) + \sum_{i=1}^{m} - logP(v_{i}^1|w, v_{0:i-1}^1) \\ 
    & + ... + \sum_{i=1}^{m} - logP(v_{i}^k|w, v^1, ..., v^{k-1}, v_{0:i-1}^k)
    \end{split}
\end{equation*}
\normalsize
where $n=T+2$ and $m=h\times w+2$, and $w_0, w_n, v_{0}^{1}, v_{m}^{1},$ $\ldots, v_{0}^{k}, v_{m}^{k}$ are special tokens.

At inference, for generating an X-ray of a specific view, the input to the model is $[\wb;\vb^1;...;\vb^{k-1}]$, meaning that the report embeddings are followed by X-ray embeddings of other views (if available for this study.).

\section{Experiments}
\subsection{Dataset}
MIMIC-CXR \citep{johnson2019mimic} contains 377,110 chest X-rays from 227,835 radiology studies.
Each study has one or multiple chest X-rays and a single report.
We select a total of 208,534 studies that contain at most 3 chest X-rays composed of the most common views, namely PA, AP, and LATERAL\footnote{A study can have PA, PA, LAT or PA, LAT, or just AP.}.
Appendix ~\ref{apd:dset_statistic} shows the statistics of chest X-ray view composition in each study.
From the report, we use the two primary sections, namely Findings and Impression.
We follow the official split of MIMIC-CXR (train 204,102, valid 1,659 test 2,773).

\subsection{Evaluation Metrics}
We evaluate the generated chest X-rays in various aspects, from sample quality to clinical efficacy.
FID is the standard evaluation metrics in generative models, but it is not appropriate to capture complex medical concepts.
Therefore, we use an additional metric, including 14-diagnosis classification
We also perform human evaluation.

\subsubsection{Statistical Evaluation}
For FID \citep{heusel2017gans}, we compute the distances of feature statistics between the original X-rays from the test set and the generated X-rays with the 1024-dimensional feature of the DenseNet-121 pretrained on chest X-ray datasets \citep{cohen2022torchxrayvision}.
\subsubsection{Clinical Efficacy Evaluation} \label{sec:clinical_efficacy_evalutions}
For 14-diagnosis classification, We train DenseNet-121 with positive labels extracted from the Findings and Impression sections using CheXpert labeler \citep{irvin2019chexpert}.
The model then predicts the classes of the generated chest X-rays.
We report micro-averaged AUROC.

\subsubsection{Human Evaluation}
Using 100 triples of an original chest X-ray, a generated chest X-ray from our model, and a baseline, we ask three board-certified clinicians to evaluate each chest X-ray on three aspects: (1) realism, (2) alignment with the given report, and (3) the view position among PA, AP, and LATERAL views.
Both (1) and (2) are rated on a scale from 1 (worst) to 5 (best).
The triples consist of 33 triples from PA and AP and 34 triples from LATERAL.
The clinicians consist of two radiologists and one neurosurgeon, and the X-rays are presented in random order for each triple.

\subsection{Experiment Design}
\label{sec:experiment_design}

\subsubsection{The Effect of Multi-view Chest X-rays}
To evaluate the effect of using multi-view chest X-rays on the generation quality, we divide the test dataset into three groups based on the number of chest X-rays per study.
These groups include studies with one X-ray (S w/1), two X-rays (S w/2), and three X-rays (S w/3).
We evaluate our model by incrementally increasing the number of input chest X-rays within each group.
For example, in the group of studies with two X-rays (S w/2), we first only use the report as the input condition for chest X-ray generation.
Next, we use both the report and the remaining chest X-ray as the input condition.
Then we compare the generated chest X-rays under these different conditions.

\subsubsection{The Ability to Generate Specific Views}
We evaluate the impact of the special tokens in generating specific views by asking the three clinicians to identify the view positions of the generated chest X-rays.

\subsubsection{Comparison with Fine-tuned Stable Diffusion}
We compare \model with a fine-tuned Stable Diffusion for chest X-ray generation as proposed in Chambon et al. \citep{chambon2022roentgen}.
While various chest X-ray generation models have been proposed, only Chambon et al. \citep{chambon2022roentgen} utilize radiology reports as an input condition.
In addition, Stable Diffusion has shown great performance in image generation.

\subsubsection{Comparison with a retrieval-based approach}
Besides generating chest X-rays from reports and additional inputs, it is also possible to retrieve chest X-rays that closely match the contents of these reports. We qualitatively compare images \textit{X}, generated by \model using reports \textit{R} and additional inputs, with images \textit{X*} retrieved based on their pairing with the most similar reports \textit{R*} in the training set. This similarity is determined by the MedViLL approach \citep{moon2022multi}, which identifies \textit{R*} as the most similar report sharing exactly matching 14 disease labels with \textit{R}.

\subsubsection{The Advantage of the Unified Model}
We evaluate the advantage of a unified model compared to separate models for multi-view chest X-ray generation.
There are three variants: 1) Single$_{AP}$, 2) Single$_{PA}$, 3) Single$_{LAT.}$, where each are trained to generate only the AP view, PA view, and the Lateral view images, respectively.


\subsubsection{The Possibility for Radiology Report Generation}
Due to our model's simple architectural design, it exhibits the capability to generate radiology reports in addition to chest X-rays.
However, it is important to note that our primary focus and contributions lie in the generation of high-quality and view-specific chest X-rays.
The generation of radiology reports serves as a proof of concept, demonstrating the versatility of our model.
In Appendix ~\ref{apd:rrg}, we provide results showcasing the feasibility of generating radiology reports by swapping the order of text and image tokens in the input sequence.

\begin{table*}[hbt!]
  \vspace{-5mm}
  \caption{
  Evaluations of generated chest X-rays using FID and 14-diagnosis classification to quantify the effect of using multi-view chest X-rays in chest X-ray generation.
  src., tar., and LAT. are short for source, target, and LATERAL, respectively.
  In each group, best values are emboldened and second-best underlined.
    }
    {\begin{adjustbox}{width=\textwidth, center}
    \begin{tabular}{cccccccccccc}
    \toprule
    \multirow{2}{*}{Group}  &
    \multirow{2}{*}{Input}  &
    \multirow{2}{*}{(src. $\rightarrow$ tar.)}  &
    \multicolumn{4}{c}{FID ($\downarrow$)}  &
    \multicolumn{4}{c}{micro AUROC}\\
    \cmidrule(lr){4-7}
    \cmidrule(lr){8-11}
    & & & ALL & AP & PA & LAT. & ALL & AP & PA & LAT. \\

    \midrule
    \textbf{S w/1} &	1 of 1 & ($\wb \rightarrow \vb^1$) & \makecell{ 25.86 \\ (25.727, 25.993) }	& \makecell{ 25.986 \\ (25.858, 26.113) } & \makecell{ 74.189 \\ (73.425, 74.953) } & \makecell{ 41.322 \\ (40.965, 41.679) } &	\makecell{ 0.747 \\ (0.747, 0.747) }	&	\makecell{ 0.751 \\ (0.75, 0.751) }	&	\makecell{ 0.756 \\ (0.751, 0.76) }	&	\makecell{ 0.565 \\ (0.562, 0.569) } \\

    \midrule
    & 1 of 2 & ($\wb \rightarrow \vb^1$)  & \makecell{ 16.965 \\ (16.916, 17.013) } & \makecell{ 26.878 \\ (26.699, 27.058) } & \makecell{ 17.778 \\ (17.696, 17.859) } & \makecell{ 20.947 \\ (20.872, 21.021) } &	\makecell{ 0.664 \\ (0.663, 0.664) }	&	\makecell{ 0.74 \\ (0.739, 0.741) }	&	\makecell{ 0.642 \\ (0.641, 0.643) }	&	\makecell{ 0.634 \\ (0.633, 0.635) } \\

    \cmidrule(lr){2-11}
    \textbf{S w/2} &	2 of 2 & ($\wb, \vb^2 \rightarrow \vb^1$)  & \makecell{ \textbf{9.186} \\ (9.133, 9.239) }	&	\makecell{ \textbf{22.071} \\ (21.827, 22.316) }	&	\makecell{ \textbf{8.337} \\ (8.301, 8.373) }	&	\makecell{ \textbf{9.088} \\ (9.054, 9.122) } &	\makecell{ \textbf{0.712} \\ (0.712, 0.713) }	&	\makecell{ \textbf{0.753} \\ (0.752, 0.754) }	&	\makecell{ \textbf{0.692} \\ (0.691, 0.693) }	&	\makecell{ \textbf{0.702} \\ (0.701, 0.702) } \\
        
    \cmidrule(lr){2-11}
    & \makecell{Diff. \\ (2of2 - 1of2)} & - 
    & \makecell{-7.779 \\(-7.851, -7.707)} & \makecell{-4.807 \\ (-5.110, -4.504)} & \makecell{-9.441 \\ (-9.530, -9.351)} & \makecell{-11.858 \\ (-11.941, -11.776)}
    & \makecell{0.049 \\ (0.048, 0.049)} & \makecell{0.013 \\ (0.012, 0.014)} & \makecell{0.050 \\ (0.049, 0.051)} & \makecell{0.067 \\ (0.066, 0.068)}\\

    \midrule
    &	1 of 3 & ($\wb \rightarrow \vb^1$) & \makecell{ 21.148 \\ (21.049, 21.246) } & \makecell{ 39.049 \\ (38.714, 39.383) } & \makecell{ 27.051 \\ (26.778, 27.325) } & \makecell{ 24.846 \\ (24.699, 24.992) } &	\makecell{ 0.668 \\ (0.667, 0.669) }	&	\makecell{ \underline{0.711} \\ (0.709, 0.713) }	&	\makecell{ 0.666 \\ (0.664, 0.668) }	&	\makecell{ 0.643 \\ (0.642, 0.644) } \\

    \cmidrule(lr){2-11}
     &	2 of 3 & ($\wb, \vb^2 \rightarrow \vb^1$) & \makecell{ \underline{12.792} \\ (12.698, 12.887) }	&	\makecell{ \underline{23.912} \\ (23.524, 24.299) }&	\makecell{ \underline{14.606} \\ (14.381, 14.83) }	&	\makecell{ \underline{16.778} \\ (16.677, 16.878) } &	\makecell{ \underline{0.694} \\ (0.693, 0.695) }	&	\makecell{ 0.689 \\ (0.687, 0.691) }	&	\makecell{ \textbf{0.717} \\ (0.716, 0.719) }&	\makecell{ \textbf{0.679} \\ (0.678, 0.681) } \\

    \cmidrule(lr){2-11}
    \textbf{S w/3}  & 3 of 3& ($\wb, \vb^2, \vb^3 \rightarrow \vb^1$) & \makecell{ \textbf{12.684} \\ (12.588, 12.781) }	&	\makecell{ \textbf{23.695} \\ (23.361, 24.03) }&	\makecell{ \textbf{14.517} \\ (14.285, 14.75) }	&	\makecell{ \textbf{16.499} \\ (16.403, 16.595) } &	\makecell{ \textbf{0.699} \\ (0.698, 0.7) }	&	\makecell{ \textbf{0.72} \\ (0.718, 0.722) }	&	\makecell{ \underline{0.716} \\ (0.714, 0.717) }	&	\makecell{ \underline{0.675} \\ (0.673, 0.676) } \\

    \cmidrule(lr){2-11}
    & \makecell{Diff. \\ (3of3 - 1of3)} & - & \makecell{-8.4631 \\ (-8.6264, -8.2997)} & \makecell{-15.3531 \\ (-15.9502, -14.756)} & \makecell{-12.5341 \\ (-12.9476, -12.1205)} & \makecell{-8.3463 \\ (-8.5437, -8.149)} & \makecell{0.0319 \\ (0.0302, 0.0335)} & \makecell{0.0088 \\ (0.0054, 0.0122)} & \makecell{0.0498 \\ (0.0469, 0.0528)} & \makecell{0.0318 \\ (0.0294, 0.0342)}\\
    
    & \makecell{Diff. \\ (3of3 - 2of3)} & - & \makecell{-0.1078 \\ (-0.2712, 0.0555)} & \makecell{-0.216 \\ (-0.8131, 0.381)} & \makecell{-0.0881 \\ (-0.5017, 0.3254)} & \makecell{-0.2783 \\ (-0.4756, -0.081)} & \makecell{0.0055 \\ (0.0038, 0.0071)} & \makecell{0.0311 \\ (0.0277, 0.0345)} & \makecell{-0.0016 \\ (-0.0045, 0.0013)} & \makecell{-0.0046 \\ (-0.007, -0.0022)}\\
    
    & \makecell{Diff. \\ (2of3 - 1of3)} & - & \makecell{-8.3553 \\ (-8.5186, -8.1919)} & \makecell{-15.137 \\ (-15.7341, -14.54)} & \makecell{-12.4459 \\ (-12.8595, -12.0324)} & \makecell{-8.068 \\ (-8.2654, -7.8707)} & \makecell{0.0264 \\ (0.0248, 0.028)} & \makecell{-0.0223 \\ (-0.0257, -0.0189)} & \makecell{0.0514 \\ (0.0485, 0.0544)} & \makecell{0.0364 \\ (0.034, 0.0388)}\\
    
    \bottomrule
    \end{tabular}
    \end{adjustbox}
    }
\label{tab:cxr_fid_classification_multiview}
\end{table*}

\section{Results and Discussion}
The statistical significance is determined by calculating the confidence interval for the difference between the two group means.
A 95\% confidence interval ($\alpha$ = 0.05) is obtained by performing a non-parametric bootstrap. 1,000 bootstrap samples of the same size as the original test dataset are randomly taken from the dataset with replacement.
In each table, numbers within parentheses indicate 95\% CI.
Diff.() indicates the confidence interval for the difference between the two means.
Additionally, as the lower FID score indicates better performance, the negative mean FID difference reflects better performance.

\subsection{The Effect of Multi-view Chest X-rays}
We investigate the effect of inputting multi-view chest X-rays on the generation ability.
As described in Section~\ref{sec:experiment_design}, we divide test dataset into three groups (S w/1, w/2, and w/3) and evaluate within each group.

For chest X-ray generation, we use the report as the input condition and also incrementally add the rest of the chest X-rays as input.
Table~\ref{tab:cxr_fid_classification_multiview} shows FID and 14-diagnosis classification results, respectively.
In the ALL view of the S w/2 group, we can observe that \textit{2 of 2} achieves significantly higher performance than \textit{1 of 2} in both statistical (FID) and clinical efficacy (AUROC: \textit{2of2} – \textit{1of2} = 0.049, [95\% CI 0.048, 0.049]) metrics. Also, \textit{2 of 2} significantly outperforms \textit{1 of 2} in the individual views (AP, PA and Lateral).
In the ALL view of S w/3 group, using additional chest X-rays (\textit{2 of 3} and \textit{3 of 3}) shows significantly higher performance when compared to using only the report (\textit{1 of 3}) across all metrics (AUROC: \textit{3of3} – \textit{1of3} = 0.032, [95\% CI 0.030, 0.034] and \textit{2of3} – \textit{1of3} = 0.026, [95\% CI 0.025, 0.028]). In the PA and Lateral views, both \textit{2 of 3} and \textit{3 of 3} significantly outperform \textit{1 of 3} across all metrics. As for the AP view, on the other hand, although both \textit{2 of 3} and \textit{3 of 3} show significantly lower FID (the lower the better) than \textit{1 of 3}, \textit{2 of 3} does not show significantly superior 14-diagnosis classification performance than \textit{1 of 3}. We believe this is partly due to the small number of AP views in the S w/3 group (refer to Table~\ref{tab:dataset_view_composition} more details), which also could be the cause for generally higher FID scores.
Moreover, note that \textit{3 of 3} does not always outperform \textit{2 of 3} in some metrics. Specifically, the AP view and the PA view do not show statistically significant differences between \textit{3 of 3} and \textit{2 of 3} in terms FID. Also, \textit{3 of 3} has significantly lower AUROC performance than \textit{2 of 3} in the Lateral view (mean AUROC Lateral difference –0.005, [95\% CI –0.007, -0.002]). We believe this is because the studies with three chest X-rays account for only a small percentage of the entire train dataset (8.5\%, refer to Table~\ref{tab:dataset_view_composition} for more details.). Therefore, there is less opportunity for the model to learn the \textit{3 of 3} input format during training. We can conclude that utilizing multiple X-ray views as input generally helps the model generate more accurate chest X-rays that can capture the abnormal findings in the report and other chest X-rays.

\begin{table*}[hbt!]
  \caption{
  Human evaluation Average means and standard deviations across three clinicians.
  }
  \vspace{-3mm}
    {
    \begin{adjustbox}{width=\textwidth} 
    \begin{tabular}{cccccccccccc}
    \toprule
    \multirow{2}{*}{\bfseries Models}  &
    \multicolumn{4}{c}{\bfseries Realism}  &
    \multicolumn{4}{c}{\bfseries Alignment}  &
    \multicolumn{3}{c}{\bfseries View Position} \\
    \cmidrule(lr){2-5}
    \cmidrule(lr){6-9}
    \cmidrule(lr){10-12}
    & \bfseries ALL & \bfseries AP & \bfseries PA & \bfseries LATERAL & \bfseries ALL & \bfseries AP & \bfseries PA & \bfseries LATERAL & \bfseries AP & \bfseries PA & \bfseries LATERAL \\
    \midrule
     Original Image & 4.177 $\pm$ 0.793 & 4.294 $\pm$ 0.703 & 4.281 $\pm$ 0.579 & 3.961 $\pm$ 0.912 & 3.977 $\pm$ 1.002 & 4.196 $\pm$ 0.855 & 4.156 $\pm$ 0.793 & 3.588 $\pm$ 1.123 & 0.843 $\pm$ 0.632 & 0.583 $\pm$ 0.487 & 1.0 $\pm$ 0.0 \\
     \model & 4.193 $\pm$ 0.675 & 4.206  $\pm$ 0.659 & 4.188 $\pm$ 0.626 & 4.186 $\pm$ 0.674 & 3.583 $\pm$ 1.013 & 3.559 $\pm$ 1.028 & 3.719 $\pm$ 0.928 & 3.48 $\pm$ 1.043 & 0.755 $\pm$ 0.415  & 0.667 $\pm$ 0.461  & 1.0 $\pm$ 0.0 \\
     Stable Diffusion & 2.09 $\pm$ 0.951 & - & - & - & 1.827 $\pm$ 0.812 & - & - & - & - & - & -\\
    \bottomrule
    \end{tabular}
    \end{adjustbox}}
    \vspace{-5mm}
\label{tab:human_eval}
\end{table*}

A key finding in this experiment is that considering the relations between the multi-view chest X-rays of the same study is important, as they provide valuable information.
We observe that using multi-view chest X-rays can faithfully capture abnormal findings in chest X-ray generation, as \textit{2 of 2} and \textit{3 of 3} show statistically significant differences compared to \textit{1 of 2} and \textit{1 of 3} input formats.
Although \textit{2 of 3} sometimes demonstrates inferior performance than \textit{1 of 3} on clinical efficacy metrics (AUROC of the AP view), the overall performance demonstrated by the ALL view suggests the effectiveness of utilizing more information rather than less information.

\subsection{The Ability to Generate Specific Views}
View Position column in Table~\ref{tab:human_eval} confirms that the view-specific special tokens can capture refined features of each view.
Specifically, the lateral view result (Lateral: Original 1.0 vs \model 1.0) shows that the view-specific special tokens can properly capture the characteristics of the lateral view that are distinct from the frontal view.
In addition, the 14-disease classification results in Table~\ref{tab:cxr_fid_classification_multiview} support that our model does not simply generate the lateral appearance of the chest but generates the lateral chest X-rays that faithfully reflect the abnormal findings. 
The generated AP view images are certainly distinguishable from PA view images, but not as clearly as the original AP view images (AP: Original 0.843 vs \model 0.755), indicating that the AP view special tokens do not perfectly capture the characteristics of the AP view. On the other hand, given that the generated PA view images are more distinguishable than the original PA view images (PA: Original 0.583 vs \model 0.667), we can infer that the PA view special tokens are already capturing the characteristics of the PA view as best as possible.
These results suggest that the view-specific special tokens are effective in generating chest X-rays in specific views, and that our model can even generate the desired views even if they do not exist in reality.
The green dashed boxes in Fig.~\ref{fig:gen_cxr} show the generated chest X-rays that do not exist in the study.
We can observe that the generated absent views have anatomical similarities to other existing views within the same study.

\begin{table}[hbt!]
  \caption{
  Comparison of \model and the fine-tuned Stable Diffusion for chest X-ray generation.
  }
  \vspace{-3mm}
    {
    \begin{adjustbox}{width=\columnwidth} 
    \begin{tabular}{ccc}
    \toprule
    Models & FID ($\downarrow$) & micro AUROC \\
    \midrule
    Stable Diffusion (S.D) &
    78.965 (78.883, 79.046) & 
    0.589 (0.589, 0.589) \\
    
    \model &
    \textbf{19.212} (19.157, 19.267) &
    \textbf{0.711} (0.711, 0.711)\\
    
    \makecell{Diff. \\ (\model $-$ S.D)} & -59.753 (-59.852, -59.655) &
    0.122 (0.122, 0.122)\\
    
    \bottomrule
    \end{tabular}
    \end{adjustbox}
    }
    \vspace{-5mm}
\label{tab:stable_diffusion}
\end{table}

\subsection{Comparison with Stable Diffusion}
Table~\ref{tab:stable_diffusion} shows the chest X-ray generation performances of \model and the fine-tuned Stable Diffusion.
For a fair comparison, our model generates chest X-rays using only radiology reports as input, without inputting any additional chest X-rays (i.e. \model uses \textit{1 of 1}, \textit{1 of 2}, and \textit{1 of 3}, respectively from S w/1, S w/2, and S w/3). We can observe that \model significantly outperforms the fine-tuned Stable Diffusion across all metrics (mean AUROC difference 0.122, [95\% CI 0.122, 0.122]).
We believe that these performance differences mainly arise from the drastic difference in pixel distributions between the chest X-ray images and the general domain images used for originally training Stable Diffusion, and the difference in the length of input text (i.e. long radiology report VS short image captions). 
In addition, our model proves again that using additional chest X-rays can effectively generate more realistic and accurate chest X-rays when comparing Tables~\ref{tab:cxr_fid_classification_single} and~\ref{tab:stable_diffusion}, with 14-diagnosis classification AUROC of 0.728 [95\% CI 0.728, 0.729] VS 0.711 [95\% CI 0.711, 0.711].

\subsection{Human Evaluation}
Table~\ref{tab:human_eval} confirms that \model can generate realistic chest X-rays comparable to the original.
More specifically, the generated frontal X-rays score 0.091 points lower than the original image (Original 4.288 vs \model 4.197 on a 1-5 scale).
One of the reasons of this difference can be the fact that the lines and tubes are sometimes generated in the wrong positions, and the details of the supporting device is insufficiently depicted.
In terms of alignment, both the original and \model attain less than 4 points for the lateral view.
This is because reports are usually written based on the frontal view, and since the lateral view plays an auxiliary role, much information cannot be found in the lateral view. Thus, focusing on the frontal view results, \model scores 0.538 points lower than the original image (Original 4.177 vs \model 3.629 on a 1-5 scale).
This difference mainly arises because our model occasionally fails to fully reflect in the X-rays the abnormalities in the report.
We can conclude that our model can generate chest X-rays similar to the original, but sometimes dose not faithfully reflect the contents in the report. Also, we can observe that the view-specific special tokens can capture refined features of each view, enabling the model to generate view-specific X-rays (AP: Original 0.843 vs \model 0.755, PA: Original 0.583 vs \model 0.667, Lateral: Original 1.0 vs \model 1.0).
In addition, our model scores higher than the baseline for both realism (\model 4.193 vs Stable Diffusion 2.09 on a 1-5 scale) and alignment (\model 3.583 vs Stable Diffusion 1.827 on a 1-5 scale).
Note that the baseline fails to learn view-specific information; thus, we do not evaluate its ability to generate images of specific views.

\subsection{The Advantage of the Unified Model}

We study the advantage of training a unified model for multi-view chest X-ray generation.

In Table~\ref{tab:cxr_fid_classification_single}, we compare our model with Single$_{AP}$, Single$_{PA}$, and Single$_{LAT.}$.
In terms of the statistical metric (FID, the lower the better), \model outperforms the single models only in the PA case.
In terms of the clinical efficacy metric (14-diagnosis classification), however, it shows significantly superior performance than all single models: \model – Single$_{AP}$ = 0.066, [95\% CI 0.065, 0.066], \model – Single$_{PA}$ = 0.007, [95\% CI 0.007, 0.008], and \model – Single$_{LAT.}$ = 0.027, [95\% CI 0.027, 0.028].
This suggests that training a model to generate multiple views helps the model to correctly capture the abnormalities described in the report.

From these results, we demonstrate that \model is comparable, if not superior, to the various single models tailored to generate only its specific modality.
Specifically, only the mean FID difference of PA outperforms the single model in the statistical metric, but except for this, \model significantly outperforms the single models across all metrics.
This suggests that our model can generate multi-view chest X-rays with clinically meaningful information.
We can conclude that bidirectional training has a synergistic effect on generation tasks and also can save time and computational costs, as opposed to training multiple single models.
\begin{table*}[hbt!]
  \vspace{-1cm}
  \caption{
  Comparison of \model with various single models to evaluate the impact of the unified model in chest X-ray generation.
  The FID scores for the original image are calculated with the same number of train set as the test set.
  Each AP, PA and LAT. column shows the performance measured by dividing the generated chest X-rays according to their original view position.
  }
    {
    \begin{adjustbox}{width=\textwidth}
    \begin{tabular}{ccccccccc}
    \toprule
    \multirow{3}{*}{Models}  &
    \multicolumn{4}{c}{FID ($\downarrow$)} &
    \multicolumn{4}{c}{micro AUROC} \\
    \cmidrule(lr){2-5}
    \cmidrule(lr){6-9}
    & ALL & AP & PA & LAT. & ALL & AP & PA & LAT. \\
    
    \midrule
    Original Image & \makecell{ 0.541 \\ (0.531, 0.551) } & \makecell{ 1.15 \\ (1.124, 1.177) }	& \makecell{ 1.611 \\ (1.59, 1.632) }	& \makecell{ 1.082 \\ (1.068, 1.096) } & \makecell{0.81 \\ (0.809, 0.81)} & \makecell{ 0.808 \\ (0.808, 0.808) }& \makecell{ 0.812 \\ (0.812, 0.812) } & \makecell{ 0.793 \\ (0.793, 0.794) } \\
    
    Single$_{AP}$  &	-	&	\makecell{ \textbf{16.172} \\ (16.037, 16.307) }	&	-	&	- & -	&	\makecell{0.689 \\ (0.689, 0.689)}	&	-	&	- \\
    
    Single$_{PA}$ &	-	&	-	&	\makecell{ 7.579 \\ (7.553, 7.605) }	&	-  & -	&	-	&	\makecell{0.697 \\(0.696, 0.697)}	&	- \\
    
    Single$_{LAT.}$  &	-	&	-	&	-	&	\makecell{ \textbf{8.242} \\ (8.222, 8.261) } & -	&	-	&	-	&	\makecell{0.667 \\ (0.667, 0.668)} \\
    
    \model &	
    \makecell{ 10.582 \\ (10.554, 10.609) }	&	
    \makecell{ 17.639 \\ (17.572, 17.705) }	&	
    \makecell{ \textbf{6.324} \\ (6.302, 6.347) }	&	
    \makecell{ 9.553 \\ (9.531, 9.575) } &
    \makecell{ 0.728 \\ (0.728, 0.729) }	&	\makecell{ \textbf{0.755} \\ (0.755, 0.755) }	&	\makecell{ \textbf{0.704} \\ (0.704, 0.705) }	&	\makecell{ \textbf{0.695} \\ (0.694, 0.695) } \\

    \makecell{Diff. \\ (\model $-$ Single$_{each view}$)} & - & \makecell{1.467 \\ (1.317, 1.618)} & \makecell{-1.255 \\ (-1.290, -1.221)} & \makecell{1.311 \\ (1.282, 1.341)} & - & \makecell{0.066 \\ (0.065, 0.066)} & \makecell{0.007 \\ (0.007, 0.008)} & \makecell{0.027 \\ (0.027, 0.028)}\\
    
    \bottomrule
    \end{tabular}
    \end{adjustbox}}
\label{tab:cxr_fid_classification_single}
\end{table*}

\subsection{Comparison with a Retrieval-based Approach}
Fig.~\ref{fig:retrieval} shows the results.
The first sample shows that the image \textit{X}, generated through our approach using the additional input view and the report \textit{R}, accurately reflects the patient's gender information.
In contrast, the retrieved image \textit{X*} paired with the report \textit{R*}, which is most similar to \textit{R}, fails to incorporate this detail.
Moreover, in identifying R*, even though disease label information was used, it did not capture the location of support devices, leading to the retrieved image X* inaccurately reflecting the precise position of support devices. This indicates the need for advanced techniques to consider all elements in retrieval effectively.
The second sample demonstrates that \textit{X*} does not account for the patient's obesity level, as this information is absent in the report.
Thus, it fails to reflect the patient's actual physical condition.
In the third sample, the report \textit{R} fails to mention a support device, yet the generated image \textit{X}, enhanced by an additional lateral view, accurately includes this detail, in contrast to the retrieved image \textit{X*} which lacks it.
Moreover, despite the absence of gender information in the report, the generated image \textit{X} correctly represents the patient's gender.
These examples illustrate the advanced capabilities of our approach to generate images that accurately include details, even those not explicitly stated or omitted in the reports. In contrast, the retrieval-based approach often fails to capture details that are not explicitly mentioned. This comparison underscores the limitations of the retrieval method in handling complex clinical scenarios effectively.

\subsection{Qualitative Examples}
\begin{figure*}[h]
    \centering
    \includegraphics[height=16cm]{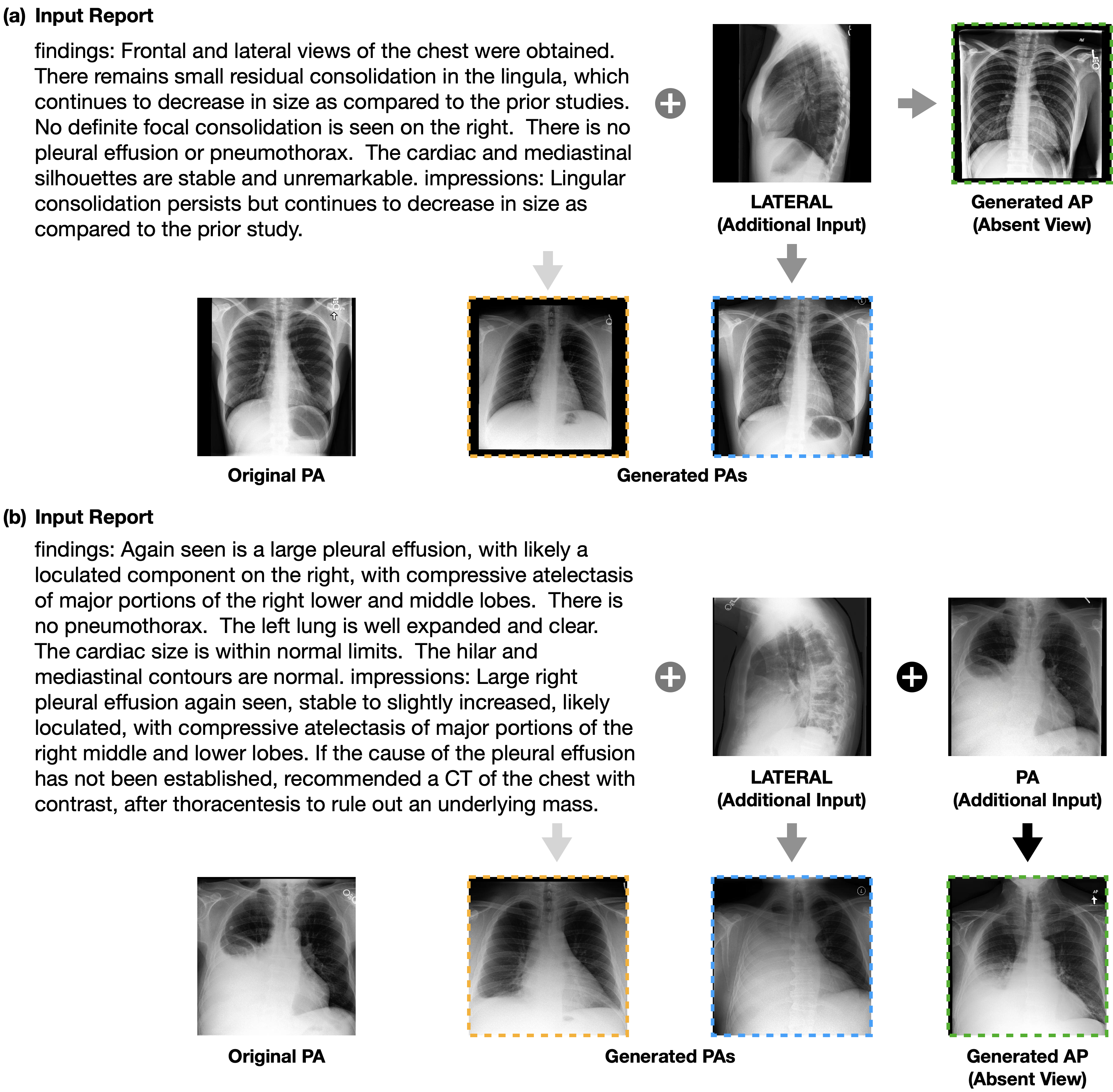}
    \vspace{-3mm}
    \caption{
    Generated chest X-rays of \model.
    (a) Based only on the report, the generated PA in the orange dashed box draws a rather small portion of the consolidation in the lingula, as is written in the report. Based on an additional lateral view, the generated PA in the blue dashed box draws a consolidation that is of more similar size as that of the original PA.
    (b) The generated PA conditioned only on the report (orange dashed box) draws relatively small-sized pleural effusion while the report says ``large right pleural effusion''. However, by adding an additional lateral view (blue dashed box), \model can properly generate the PA view with large pleural effusion.
    }
\label{fig:gen_cxr}
\end{figure*}

Fig.~\ref{fig:gen_cxr} (a) shows that \model can generate realistic chest X-ray images even when conditioned only on the report, describing a small consolidation in the lingula as described by the report.
When given an additional view, \model generates an image that is more similar to the original image, showing its ability to take advantage of both input modalities.
Fig.~\ref{fig:gen_cxr} (b), on the other hand, shows a scenario where the generated image, conditioned solely on the report, fails to accurately capture all the details described in the report.
Although the report says ``large right pleural effusion'', the generated image depicts a rather small pleural effusion.
When given an additional view, however, \model can draw pleural effusion that is of the similar size as that of the original image.
Furthermore, both figures show that the view-specific special tokens enable \model to generate the desired views, even when they do not exist in reality.
All figures are confirmed by the clinicians.

\section{Limitation and Conclusion}
Here, we propose for the first time a novel approach to generate chest X-rays with specific views, addressing the limitations of existing methods that primarily focus on generating frontal views.
Our model introduces specialized tokens and leverages multi-view information to enable users to generate chest X-rays according to their desired views.
Our approach has some limitations, each providing opportunities for future work.
First, due to the nature of the real-world patient dataset, the report often contains references to previous studies (e.g. unchanged, increase, and compared to previous radiographs). These references have the potential to impact the quality of chest X-ray generation.
In the future, we plan to use CXR-PRO\citep{ramesh2022improving}, a refined dataset that removes comparison phrases, to generate clinically accurate chest X-rays.
Second, the human evaluation confirms that our model generates chest X-rays that sometimes fail to fully reflect the facts in the given report (Original 3.977, \model 3.583 on a 1-5 scale in Table.~\ref{tab:human_eval}).
In addition, the position and shape of the support device are slightly different from the original image, so we can infer that our model sometimes has difficulty capturing fine details.
We defer addressing these challenges for the future.


\acks{
This work was supported by Samsung Electronics (No.IO201211-08109-01), Institute of Information \& Communications Technology Planning \& Evaluation (IITP) grant (No.2019-0-00075), National Research Foundation of Korea (NRF) grant (NRF-2020H1D3A2A03100945), and the Korea Health Industry Development Institute (KHIDI) grant (No.HI21C1138) funded by the
Korea government (MSIT, MOHW).
}

\clearpage
\bibliography{chil-sample}

\clearpage
\appendix

\section{Dataset Statistic}\label{apd:dset_statistic}
\begin{table*}[ht]
\vspace{-5mm}
  \caption{
  Composition of chest X-ray views in each study.
  S w/1, S w/2, and S w/3 indicate the number of chest X-rays per study.
  LAT. is short for Lateral.}
    {
    \begin{adjustbox}{width=\textwidth, center}
    \begin{tabular}{cc}
    \begin{tabular}{ccccc}
    \toprule
    \bfseries Group &
    \bfseries Split &
    \bfseries AP &
    \bfseries PA &
    \bfseries LAT. \\
    \midrule
    & Train & 	91,736 &  85 &  1,596\\
    \cmidrule{2-5}
    S w/1  & Valid & 782 & 1	& 12 \\
    \cmidrule{2-5}
    & Test & 1,428 &  3 &  29\\
    \bottomrule
    \end{tabular} & 
    
    \begin{tabular}{cccccccc}
    \toprule
    \bfseries Group &
    \bfseries Split &
    
    \bfseries (PA, LAT.) &
    \bfseries (AP, LAT.) &
    \bfseries (AP, AP) &
    \bfseries (LAT., LAT.) &
    \bfseries (PA, PA) &
    \bfseries (AP, PA) \\
    
    \midrule
    & Train & 68,600 & 13,971 & 9,853 & 471	& 315 & 105 \\
    \cmidrule{2-8}
    S w/2  & Valid & 513 & 95 & 90 & 3 & 2	& 2\\
    \cmidrule{2-8}
    & Test & 671 & 212 & 162 & 10 & 3 & 1 \\
    \bottomrule
    \end{tabular} \\
    \end{tabular}
    \end{adjustbox}}
\label{tab:dataset_view_composition}
\end{table*}

\begin{table*}[ht]
    {
    \begin{adjustbox}{height=1cm, center}
    \begin{tabular}{cccccccc}
    \toprule
    \bfseries Group &
    \bfseries Split &
    
    \bfseries (PA, PA, LAT.) & 
    \bfseries (AP, LAT., LAT.) &
    \bfseries (PA, LAT., LAT.) &
    \bfseries (AP, AP, LAT.) &
    \bfseries (AP, AP, AP) &
    
    \bfseries Etc. \\
    \midrule
    & Train & 8,056 &	3,968 &	3,539 &	848	& 748 & 211\\
    \cmidrule{2-8}
    S w/3  & Valid & 66 & 36 & 36 & 9 & 7 & 5\\
    \cmidrule{2-8}
    & Test & 82	& 89 & 52 & 11 & 14 & 6\\
    \bottomrule
    \end{tabular}
    \end{adjustbox}}
\end{table*}
Table~\ref{tab:dataset_view_composition} represents the compositions of chest X-ray views in each study. 

\section{Implementation Details}\label{apd:details}
\subsection{Image tokenizer}
We adopt VQ-GAN with $d_z$=256 and a codebook size of 1024.
The input image of size $512 \times 512$ is quantized into $32 \times 32 = 1024$ discrete visual tokens.
The model is trained for 540k steps with a batch size of 8, a learning rate of 4.5e-6 with the Adam optimizer.

\subsection{Text tokenizer}
We train a byte-level BPE tokenizer \citep{wang2020neural} with a minimum frequency of 2 on reports converted to lowercase.
We then obtain 14,526 unique tokens, including three special tokens $[SOS]$, $[EOS]$, $[TXT\_PAD]$.

\subsection{\model}
We set the length of word tokens $n$=256 and visual tokens $m$=1,026, including special tokens.
In this work, \model takes up to three chest X-rays as input, as the majority of studies in the MIMIC-CXR dataset have three or fewer images. However, it is able to take more images if they are available.
Our model is built on the Transformer architecture with generalized attention \citep{choromanski2020rethinking}.
The model has 12 layers, 12 heads, and 768 dimensions.
We incorporate seven special tokens (in addition to three text special tokens), namely $[SOS_{AP}]$, $[EOS_{AP}]$, $[SOS_{PA}]$, $[EOS_{PA}]$, $[SOS_{LAT}]$, $[EOS_{LAT}]$, $[IMG\_PAD]$.
Thus, the size of visual embedding function (\textit{i.e.} lookup matrix) is $f_{VE}(\cdot) \in \mathbb{R}^{N \times d}$, where $N$ = 1024 + 7, $d$ = 768, and word embedding function is $f_{WE}(\cdot) \in \mathbb{R}^{M \times d}$, where $M$ = 14,526, $d$ = 768.
We train the model for 337k steps with a batch size of 48 using four NVIDIA RTX A6000 GPUs.
We use the AdamW optimizer with a learning rate of 1.7e-4, $\beta_1$=0.9, $\beta_2$=0.999, $e=1e-8$, a weight decay of $1e-2$, and a cosine decay schedule.
We generate all samples with Top-$p$ sampling \citep{holtzman2019curious} with $p$=0.9 and temperature=0.7.

\subsection{Finetuned Stable Diffusion}
Following Chambon et al. \citep{chambon2022roentgen}, we replace the CLIP text encoder with SapBERT \citep{liu2020self} to handle both Findings and Impression sections (the CLIP tokenizer is limited to 77 tokens) and keep frozen the text encoder and VAE and only train U-Net from scratch.

\section{Radiology Report Generation}\label{apd:rrg}
\subsection{Evaluation Metrics}
We evaluate the generated reports  using metrics such as BLEU and CheXpert F1 score.
For BLEU \citep{papineni2002bleu}, we report BLEU-4 between the original and the generated reports.
For CheXpert F1 score, We extracted diagnosis labels from the original and generated reports with the CheXpert labeler. We then compare these labels and measure micro-averaged F1.

\subsection{The Effect of Multi-view Chest X-rays}
Table~\ref{tab:report_bleu_labeler_multiview} shows the effect of using multi-view chest X-rays in the radiology report generation.
We increase the input chest X-rays to generate the target report.
In the S w/2 group, although \textit{2 of 2} shows significantly lower performance than \textit{1 of 2} in terms of the simple statistical metric (BLEU-4), \textit{2 of 2} significantly outperforms \textit{1 of 2} in the clinical efficacy metrics (mean CheXpert F1 difference, 0.007 [95\% CI 0.006, 0.007]).
In the S w/3 group, \textit{3 of 3} performs significantly higher across all metrics.
These results show that using multi-view chest X-rays encourages the model to generate more clinically precise reports.
In particular, the use of multi-view chest X-rays in radiology report generation can be considered to follow the writing behavior of radiologists given that \textit{2 of 2} and \textit{3 of 3} show significantly superior performance than other input formats in clinical efficacy metric (CheXpert F1).
\begin{table}[hbt!]
  \caption{
  Evaluations of generated reports using BLEU and CheXpert F1 to quantify the effect of using multi-view chest X-rays on radiology report generation.
  src. is short for source, and tar. for target.
  Numbers within parentheses indicate 95\% CI.
  Diff.() indicates the confidence interval for the difference between the two means.
  }
    {
    \begin{adjustbox}{width=\columnwidth, center}
    \begin{tabular}{ccccc}
    \toprule
    Group & Input & (src. $\rightarrow$ tar.) & BLEU-4 & CheXpert F1 \\
    \midrule
    S w/1 & 1 of 1 & ($\vb^1 \rightarrow \wb$) & \makecell{0.042 \\ (0.042,0.042)} & 
    \makecell{0.412 \\ (0.412, 0.412)}\\
    
    \midrule
    & 1 of 2 & ($\vb^1 \rightarrow \wb$) &
    \makecell{0.056 \\ (0.056, 0.057)} & 
    \makecell{0.415 \\ (0.415, 0.415)}\\
    
    S w/2 & 2 of 2 & ($\vb^1, \vb^2 \rightarrow \wb$) & \makecell{\textbf{0.056} \\ (0.056, 0.056)} & \makecell{\textbf{0.422} \\ (0.421, 0.422)}\\
    
    & Diff. (2of2 $-$ 1of2) & - & 
    \makecell{-0.001 \\ (-0.001, -0.001)} & 
    \makecell{0.007 \\ (0.006, 0.007)}\\
    
    \midrule
    & 1 of 3 &($\vb^1 \rightarrow \wb$) & 
    \makecell{0.054 \\ (0.054, 0.054)}  & 
    \makecell{0.435 \\ (0.435, 0.436)}\\
    S w/3 & 2 of 3 & ($\vb^1, \vb^2 \rightarrow \wb$) & \makecell{\underline{0.060} \\ (0.060, 0.061)}  & 
    \makecell{\underline{0.436} \\ (0.435, 0.437)} \\
    & 3 of 3 & ($\vb^1, \vb^2, \vb^3 \rightarrow \wb$) & \makecell{\textbf{0.063} \\ (0.063, 0.063)} & \makecell{\textbf{0.451} \\ (0.450, 0.452)} \\

    & Diff. (3of3 - 1of3) & - & 
    \makecell{0.009 \\ (0.008, 0.009)} & 
    \makecell{0.019 \\ (0.014, 0.017)} \\
    & Diff. (3of3 - 2of3) & - & 
    \makecell{0.003 \\ (0.002, 0.003)} & 
    \makecell{0.016 \\ (0.014, 0.017)}\\
    & Diff. (2of3 - 1of3) & - & 
    \makecell{0.006 \\ (0.006, 0.007)} & 
    \makecell{0.0003 \\ (-0.001, 0.002)}\\
    \bottomrule
    \end{tabular}
    \end{adjustbox}
    }
\label{tab:report_bleu_labeler_multiview}
\end{table}

\subsection{The Advantage of the Unified Model}
As shown in Table~\ref{tab:report_bleu_labeler_single}, we compare our model with Single$_{report}$.
We can observe that \model significantly outperforms Single$_{report}$ in both statistical and clinical efficacy metrics (mean CheXpert F1 difference = 0.067, [95\% CI 0.066, 0.067]). This indicates that combining chest X-ray image generation as a target can effectively capture local regions that encourage the model to generate more precise reports containing abnormal findings.
\begin{table}[hbt!]
  \caption{
  Comparison of \model with a single model to evaluate the impact of the unified model in radiology report generation.
  Numbers within parentheses indicate 95\% CI.
  Diff.() indicates the confidence interval for the difference between the two means.
  }
    {
    \begin{adjustbox}{height=1.3cm, center}
    \begin{tabular}{ccc}
    \toprule
    Models & BLEU-4 & CheXpert F1 \\
    \midrule
    Single$_{report}$ & \makecell{0.038 \\(0.038 0.038)} & \makecell{0.353 \\ (0.353, 0.353)}\\
    
    \model & \makecell{\textbf{0.050} \\ (0.050 0.050)} & \makecell{\textbf{0.420} \\ (0.420, 0.420)} \\

    \makecell{Diff. \\ (\model $-$ Single$_{report}$)} & \makecell{0.012 \\ (0.012, 0.012)} & \makecell{0.067 \\ (0.066, 0.067)}\\
    \bottomrule
    \end{tabular}
    \end{adjustbox}}
    \vspace{-5mm}
\label{tab:report_bleu_labeler_single}
\end{table}

\subsection{Qualitative Examples}
\begin{figure*}[h]
    \centering
    \includegraphics[width=\textwidth * 2/3]{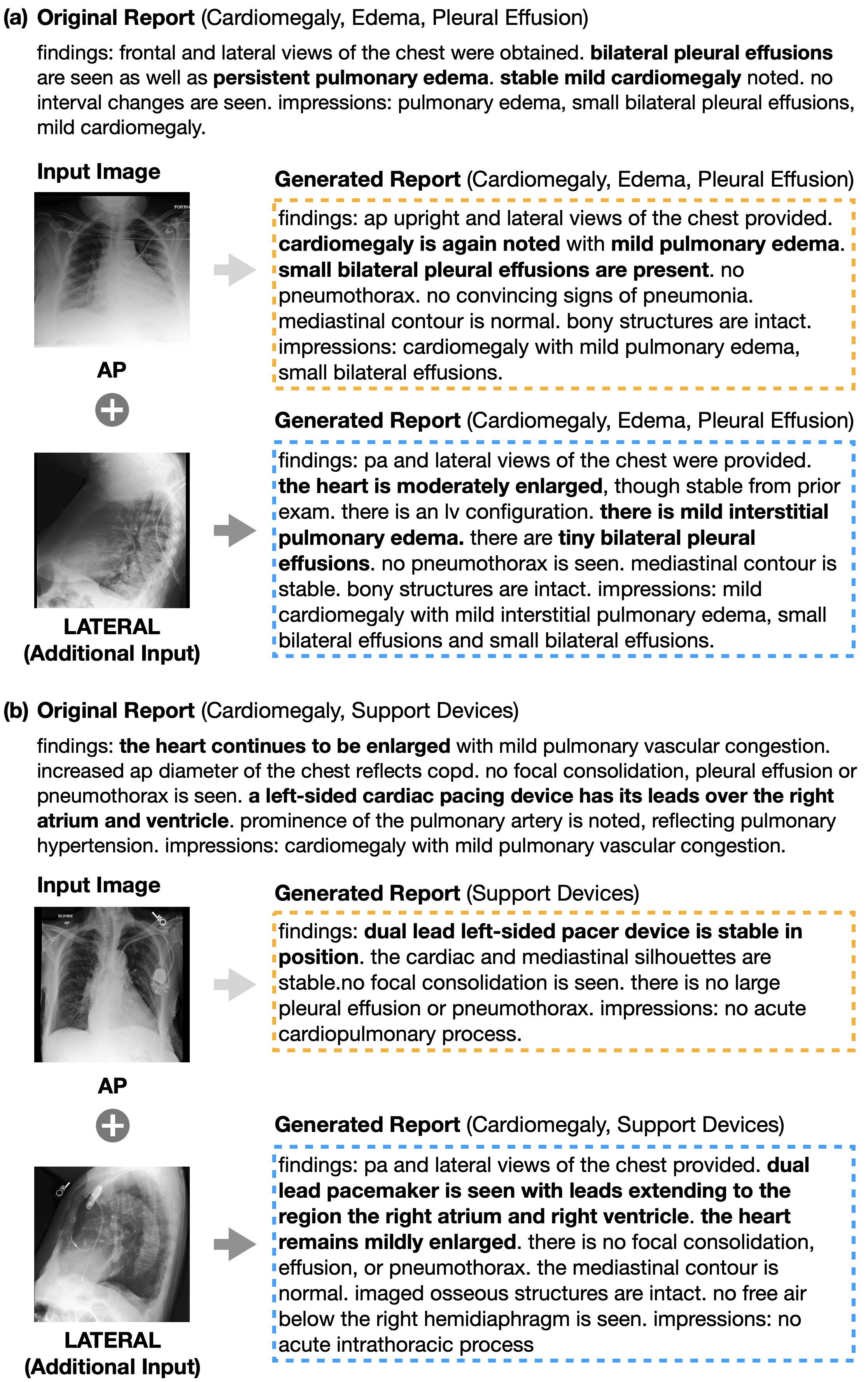}
    \caption{
    Generated radiology reports of \model.
    (a) Regardless of the number of chest X-rays input, \model can generate accurate radiology reports covering all diseases mentioned in the original report. (b) The generated report only from a single chest X-ray (orange dashed box) cannot fully capture the abnormalities in the given X-ray. With an additional chest X-ray, \model can generate a more precise report (blue dashed box) containing all diseases as described in the original report.
    }
    \vspace{-5mm}
\label{fig:gen_report}
\end{figure*}

Fig.~\ref{fig:gen_report} (a) shows an example where \model generates accurate radiology reports when given one or two chest X-ray images. Fig.~\ref{fig:gen_report} (b) shows an example where the report generated based on only one view does not capture some findings, but additional input helps the model generate more precise reports. 
All examples are confirmed by the clinicians.

\begin{figure*}[h]
    \vspace{-10mm}
    \centering
    \includegraphics[height=20.5cm]{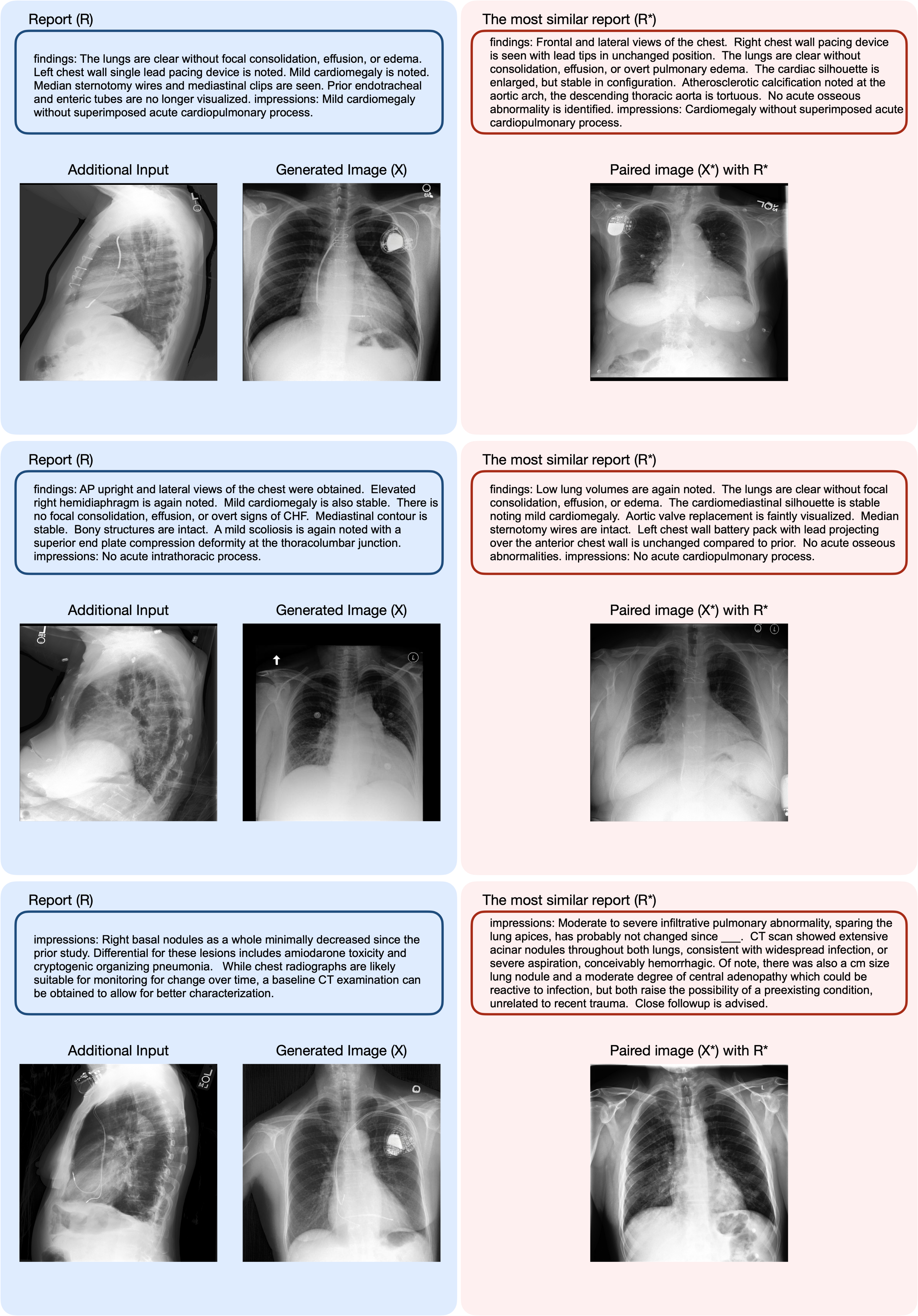}
    \caption{
    These examples highlight the advanced capabilities of our approach to generate images that accurately incorporate details, even those not explicitly stated or omitted in the reports. In contrast, they underline the limitations of a purely retrieval-based approach, which often fails to capture essential patient information such as gender or specific health conditions like obesity, especially when faced with incomplete or erroneous reports. This comparison demonstrates the inadequacy of the retrieval method in handling complex clinical scenarios.
    }
\label{fig:retrieval}
\end{figure*}

\end{document}